\documentclass[preprint,tightenlines,aps,nofootinbib,showpacs]{revtex4}
\usepackage[utf8]{inputenc}
\usepackage{amsmath}
\usepackage{amsfonts}
\usepackage{amssymb}
\usepackage{lscape} 
\usepackage{fancyhdr}
\usepackage{graphicx}
\usepackage{pifont}
\usepackage{dcolumn}
\usepackage{bm}
\usepackage{caption,subcaption}

\usepackage[none]{hyphenat}

\usepackage{latexsym}
\usepackage{graphics}
\usepackage{amssymb}
\usepackage{amsfonts}
\usepackage{mathrsfs}
\usepackage{bm}
\usepackage{amsmath}
\usepackage{color}
\usepackage[dvipsnames]{xcolor}
\usepackage{arydshln} 
\usepackage[title]{appendix}

\RequirePackage{graphicx}
\RequirePackage{mathptmx}      
\RequirePackage{flushend}
\RequirePackage[colorlinks,citecolor=blue,urlcolor=blue,linkcolor=blue]{hyperref}

\usepackage{mathtools}  
\mathtoolsset{showonlyrefs} 

\newcommand{\vecx}{\mathbf{x}}
\newcommand{\vecu}{\mathbf{u}}
\begin{document}


\title{Casimir-Lifshitz pressure on cavity walls}
\bigskip

\author{
	C. Romaniega \footnote{e-mail: cesar.romaniega@uva.es}
}

\bigskip

\affiliation{
	{Departamento de F\'isica Te\'orica, At\'omica y \'Optica,}\\ \small{Universidad de Valladolid, 47011. Valladolid, Spain.}
}

\bigskip


\begin{abstract}
We extend our previous work on the electromagnetic Casimir-Lifshitz interaction between two bodies when one is  contained within the other. We focus on the  fluctuation-induced pressure acting on the cavity wall, which is assumed to be spherical. This pressure can be positive or negative  depending on the response functions describing the bodies and the medium filling the cavity. However, we find that, under general hypotheses, the sign is independent of the geometry of the configuration. This result is based on the representation of the Casimir-Lifshitz energy in terms of transition operators. In particular, we study the components of these operators related to inside scattering amplitudes, adapting the invariant imbedding procedure to this unfamiliar scattering setup. We find that our main result is in agreement with the Dzyaloshinskii-Lifshitz-Pitaevskii result, which is obtained as a limiting case.
\end{abstract}

\maketitle

\section{Introduction}\label{sec:I}
Casimir-Lifshitz forces are a directly observable manifestation of quantum field theory. These arise from the quantum fluctuations of the electromagnetic field in the presence of material boundaries.
In the past few decades, the measurement of such forces has become increasingly precise, proving the good agreement between theory and experiment for simple geometries \cite{mohideen1998precision,chan2001quantum,munday2009measured,garrett2018measurement}.
One of the most significant achievements has been the experimental confirmation of the Dzyaloshinskii-Lifshitz-Pitaevskii (DLP) formula \cite{dzyaloshinskii1961general}.
In this case, the Casimir-Lifshitz force between two
homogeneous slabs separated by a medium  with nontrivial  electromagnetic response satisfies
\begin{equation}\label{eq:LifshitzRel}
\text{sgn}\, F_\text{int}=	-\text{sgn}[(\varepsilon_1-\varepsilon_{M})(\varepsilon_2-\varepsilon_{M})].
\end{equation}
This behaviour, hereinafter referred to as the DLP result, leads to repulsion if the permittivities of the objects $\varepsilon_{i}$ and the medium $\varepsilon_{M}$ satisfy  $\varepsilon_{i}<\varepsilon_{M}<\varepsilon_{j}$, $i,j\in\{1,2\}$.
It took almost fifty years to experimentally confirm this prediction for material bodies \cite{munday2009measured}. This was accomplished with a gold-covered sphere
and a large silica plate immersed in bromobenzene since, over a wide frequency range,   $$\varepsilon_\text{silica}<\varepsilon_\text{bromobenzene}<\varepsilon_\text{gold}.$$ 
With this, an attractive interaction was also found replacing the silica plate with a gold plate.

The experiments on Casimir force measurements have traditionally been restricted to objects like spheres and plates. Nonetheless, it is known that nontrivial  geometries modify the strength and the sign of the
 force \cite{bordag2009advances,garrett2018measurement,venkataram2020fundamental}, raising the possibility for specific applications in 
nanomechanics and nanotechnology. However, there is a lack of general theorems regarding this dependence on geometry
and boundaries.
For instance, one of the few general results on the sign of the force is for a mirror symmetric arrangement of objects \cite{kenneth2006opposites,bachas2007comment}.  In this case, the  attractive character of the force can be proved from the representation of the energy in terms of transition operators. As a result, the interaction force is also attractive between a single object and a plane  when both share  boundary conditions. This attraction may lead to the permanent adhesion of the moving parts in nanoscale machines \cite{buks2001stiction,munday2010repulsive}. In this context,
the stability of the configuration should also be considered, specially when looking for ultra-low stiction and levitating devices \cite{capasso2007casimir,munday2009measured}. Within an approach  similar to the one in \cite{kenneth2006opposites}, an extension of Earnshaw's
theorem  was stated in \cite{rahi2010constraints}. This result sets restrictive constraints on the stability of arbitrarily shaped objects  held in equilibrium by Casimir-Lifshitz forces. Nevertheless,  a possible way to avoid the assumptions of the previous no-go theorems \cite{kenneth2006opposites,rahi2010constraints}  has recently been proposed \cite{jiang2019chiral}. This is based on the introduction of a chiral medium, where  the strength of the resulting forces can be tailored in response to an external magnetic field.  

In this paper we investigate the Casimir effect in configurations in which one object lies inside the other \cite{marachevsky2001casimir,hoye2001casimir,brevik2002casimir,brevik2005casimir,dalvit2006exact,marachevsky2007casimir, zaheer2010casimir,teo2010casimir,rahi2010stable,parashar2017electromagnetic}. We extend our previous work on the pressure acting on spherical surfaces \cite{romaniega2021repulsive}. Instead of a dielectric sphere
enclosed within an arbitrarily shaped magnetodielectric body, we now consider two arbitrarily shaped objects: a  dielectric with a spherical cavity in which a magnetodielectric is enclosed. Indeed, this study is also related to \cite{cavero2021casimir}, where we study the sign of the interaction energy for scalar fields in the presence of spherical $\delta$-$\delta'$ contact interactions. In all of these cases, we have written the energy in terms of the transition operators in a similar manner as in the two above-mentioned theorems \cite{kenneth2008casimir,rahi2009scattering}. With this, in this paper we show that the signs of the interaction energy and pressure are essentially determined as in the DLP result \eqref{eq:LifshitzRel}. Furthermore, the same condition   appears in \cite{rahi2010constraints}, excluding stable equilibrium for two nonmagnetic bodies immersed in vacuum.

The paper is organized as follows. In Sec.~\ref{sec:II} we  rewrite  the interaction energy formula in terms of transition operators. This representation of the energy is  proposed to facilitate the computation of the pressure.  In Sec.~\ref{sec:InsideScatt} we  study the variation of these transition operators with respect to the radius of the cavity wall. We employ the invariant imbedding procedure for the transition matrix, which was originally developed to solve quantum mechanical scattering problems for atomic and molecular collisions. Indeed, we only need to study certain components of these operators, which are related to  the inside scattering amplitudes properly defined in this section. We also include an appendix devoted to adapt this technique for other unusual scattering setups in which the source and the detector of the scattering experiment can be inside or outside the target.
In Sec.~\ref{sec:IntPress} we prove that the sign of the interaction pressure acting on the cavity wall equals the sign of the interaction energy and that both can be written as in Eq.~\eqref{eq:LifshitzRel}.  As a consistency test, we also compare with previous results for cavity configurations.
In particular, we extend the DLP result to inhomogeneous slabs, showing the relation between the interaction pressure on the wall and the force between the slabs. We end in Sec.~\ref{sec:Exten} with some remarks on the main result and the conclusions in Sec.~\ref{conclusions}.

Throughout this paper we will use the natural units  $\hslash=c=\varepsilon_0=\mu_0=1$, neglecting fluctuations 
due to nonelectromagnetic oscillations when the medium is different from vacuum, which are usually small \cite{dzyaloshinskii1961general}.

{\section{Interaction energy}\label{sec:II}}

\begin{figure}[h!]
	\centering
	\centering\includegraphics[width=0.4\textwidth]{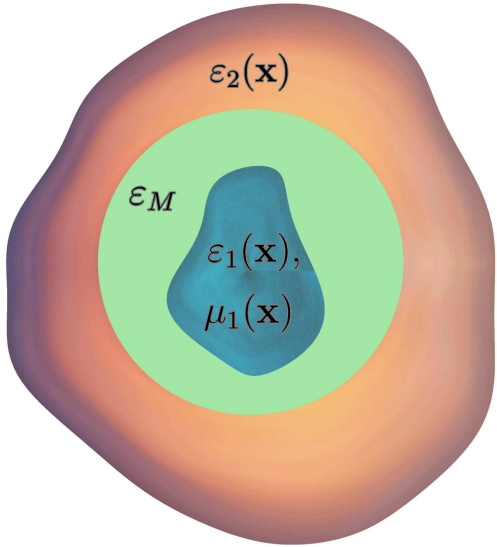}
	\caption{Cross-section view of the system under study. The pressure on the spherical cavity wall due to the interaction with the magnetodielectric object is considered. Between both objects there is a homogeneous dielectric. Each body is characterized by its corresponding permittivity $\varepsilon_{i}$ (and permeability $\mu_1$ for the inner object) being $i\in\{1,2,M\}$.\label{fig:1a}}  
\end{figure}

This section is devoted to study the interaction energy between two material bodies, one inside the other, due to the quantum fluctuations of the electromagnetic field. In particular, we consider the appropriate representation for determining the sign of the pressure acting on the  spherical cavity wall shown in Fig.~\ref{fig:1a}. 
For these bodies and the homogeneous medium filling the cavity we assume that the coupling of the  electromagnetic field to matter can be described by  continuous permittivity $\varepsilon$ and permeability ${\mu}$ functions. With this, Maxwell curl equations,
\begin{eqnarray*}
	\boldsymbol{\nabla} \times \textbf{E}(t,\mathbf{x})&=&-\partial_t\textbf{B}(t,\mathbf{x}),\\[0.5ex]
	\quad \boldsymbol{\nabla} \times \left[{\mu}^{-1}(t,\mathbf{x})\textbf{B}(t,\mathbf{x})\right]&=&-\partial_t\left[\varepsilon(t,\mathbf{x})\textbf{E}(t,\mathbf{x})\right],
\end{eqnarray*}
after Fourier transform in time, can be rearranged  to give \cite{newton2013scattering}
\begin{equation}\label{eq:SchEM}
	\left[ \boldsymbol{\nabla} \times \boldsymbol{\nabla}\times + \mathbb{V}(\omega,\mathbf{x})\right]\textbf{E}(\omega,\mathbf{x}) ={\varepsilon_{M}}\omega^2
	\textbf{E}(\omega,\mathbf{x}),
\end{equation}
where the potential  operator is
\begin{equation*}	
	\mathbb{V}(\omega,\mathbf{x})=\mathbb{I} {\omega^2} \left[\varepsilon_{M}(\omega)-\varepsilon(\omega, \textbf{x})\right] +\boldsymbol{\nabla} \times \left[\frac{1}{\mu(\omega,\textbf{x})}-\frac{1}{\mu_{M}(\omega)}\right] \boldsymbol{\nabla} \times.
\end{equation*}
Since the magnetic response of ordinary materials is typically close to one, we will focus on nonmagnetic bodies.  However, we shall see that in some cases the introduction of nontrivial permeabilities  poses no additional difficulties, especially for the inner object. Hence,
the potential operator reduces to
\begin{equation}\label{eq:Vi}
\mathbb{V}_{i}(\omega, \textbf{x}) = \mathbb{I} {V}_{i}(\omega, \textbf{x}) =\mathbb{I} \omega^2 \left[\varepsilon_{M}(\omega)-\varepsilon_{i}(\omega, \textbf{x})\right].
\end{equation}
Since $\varepsilon_{i}(\omega, \textbf{x})$ is equal to the permittivity of the medium outside each body, the spatial support of each potential satisfies $\text{supp}\,V_1\cap\,\text{supp}\,V_2=\emptyset$. 
Consequently, the Casimir-Lifshitz interaction energy between the two objects is given by the so-called \textit{TGTG} formula \cite{kenneth2008casimir,rahi2009scattering}
\begin{equation}\label{eq:ETGTG}
	E_\text{int}=\frac{1}{2\pi}\int_0^\infty \mathrm{d} \omega\,\text{Tr} \log(\mathbb{I}- {\mathbb{T}_1}\mathbb{G}^{M}_{12}\mathbb{{T}}_2\mathbb{G}^{M}_{21}).
\end{equation}
The properties of each body are encoded in its Lippmann-Schwinger $\mathbb{T}$ operator  $\mathbb{T}_{i}:\mathcal{H}_{i}\to \mathcal{H}_{i}$, being $\mathcal{H}_{i}\equiv L^2(\text{supp}\,\mathbb{V}_{i})^3$ \cite{hanson2013operator}. Furthermore, the relative position between both objects  enters through
$
\mathbb{G}^{M}_{ij}\equiv \mathbb{P}_{i}\mathbb{G}^{M} \mathbb{P}_{j}:\mathcal{H}_{j}\to \mathcal{H}_{i},
$
being $\mathbb{G}^{M}$  the  propagator across the medium and $\mathbb{P}_{i}$ the projection operator onto the Hilbert space  $\mathcal{H}_{i}$.
These electric Green's dyadics  fulfill
\begin{equation}\label{eq:DyadG}
	\left[ \boldsymbol{\nabla}\times\boldsymbol{\nabla}\times+\varepsilon_{M}(i k){k^2} \right] \mathbb{G}^{M}(ik,\mathbf{x},\mathbf{x}') =~ \mathbb{I}\delta (\mathbf{x}~ -~ \mathbf{x}'),
\end{equation}
being related to the $\mathbb{T}_{i}$ operators
of the Lippmann-Schwinger equation
\begin{eqnarray}\label{eq:LS}
\textbf{E}  &=& \textbf{E}_\text{in}- \mathbb{G}^{M} \mathbb{T}_i \textbf{E}_\text{in},\\[0.5ex]
\mathbb{T}_{i} &=& {\mathbb{V}_{i}}\left({\mathbb{I} + \mathbb{G}^{M} \mathbb{V}_{i}}\right)^{-1}.
\end{eqnarray}
Since the integral in Eq.~\eqref{eq:ETGTG} will be carried out over imaginary frequencies $\omega= i k$, we have written the analytic continuation to the imaginary frequency axis of functions such as $\varepsilon_{M}$.
The two bodies are  separated from each other so we can expand the Green's functions in terms of  free solutions  of Eq.~\eqref{eq:SchEM}. In spherical coordinates, there is a complete set of regular solutions at the origin $\{\textbf{E}_{a}^\text{reg}\}$, whose radial part is determined by the spherical Bessel function, and a complete set of  outgoing solutions   $\{\textbf{E}_{a}^\text{out}\}$, whose radial part is determined by the spherical Hankel function of the first  kind  \cite{sun2019invariant}. The subscript of these transverse solutions stands for the angular momentum values and  polarizations  $\{\ell, {m}, P\}$. Note that we have to choose the appropriate representation of $\mathbb{G}^{M}_{ij}$ depending on the relative position between both objects \cite{rahi2009scattering}. When the two bodies lie outside each other, the suitable expansion leads to the usual components of $\mathbb{T}_{i=1,2}$ in Eq.~\eqref{eq:ETGTG}, i.e.,
the scattering amplitude  related to  a  process in which a regular wave
interacts with the target and scatters outward \cite{newton2013scattering}.
However, when one body is inside the other different components of $\mathbb{T}_2$ are needed. Specifically, with the appropriate normalization, ${\mathbb{T}_1}\mathbb{G}^{M}_{12}\mathbb{{T}}_2\mathbb{G}^{M}_{21}$ can be expanded as
\begin{equation}\label{eq:ExpansionTGTG}
\sum_{a,\, b,\,c}\langle  \textbf{E}_{j}^\text{reg}, {\mathbb{T}_1} \textbf{E}_{a}^\text{reg} \rangle \langle  \textbf{E}_{a}^\text{reg}, \mathbb{G}^{M}_{12} \textbf{E}_{b}^\text{out} \rangle \langle  \textbf{E}_{b}^\text{out}, \mathbb{{T}}_2 \textbf{E}_{c}^\text{out} \rangle \langle  \textbf{E}_{c}^\text{out} , \mathbb{G}^{M}_{21} \textbf{E}_k^\text{reg} \rangle.
\end{equation}
As we shall see in Sec.~\ref{sec:InsideScatt}, $\langle  \textbf{E}_{b}^\text{out}, \mathbb{{T}}_2 \textbf{E}_{c}^\text{out} \rangle$ is related to the scattering amplitude when both the source and detector are inside the target.

Now, in order to determine the conditions leading to positive or negative values of the Casimir-Lifshitz energy, we assume that the sign of the potential $V_{i}$ in  Eq.~\eqref{eq:Vi} is constant over the whole body, being
\begin{equation}\label{eq:Signs}
	s_{i}\equiv \text{sgn}\, V_i =\pm 1 \ \ \text{if} \ \ \varepsilon_{i}(ik, \textbf{x}) \gtrless \varepsilon_{M}(i k), \ \ \forall \textbf{x}\in \text{supp}\,V_{i}.
\end{equation}
We shall see that our analysis applies to each fixed frequency so  Eq.~\eqref{eq:Signs} should hold for all of them. However,  we can simply assume constant sign over the frequencies  contributing most to the energy \cite{dzyaloshinskii1961general,munday2009measured}.
Note that from Kramers-Kronig causality conditions the permittivity satisfies $\varepsilon(ik)\geq 1$, where the equality holds for vacuum \cite{bordag2009advances}. Indeed, the vacuum Green's functions are related to the medium ones by 
$
\mathbb{G}^0(i \sqrt{\varepsilon_{M}}k,\mathbf{x},\mathbf{x}') = \mathbb{G}^{M}(i k,\mathbf{x},\mathbf{x}'),
$
being
$ \mathbb{G}^{M} $  a nonnegative operator  $\mathbb{G}^{M}\geq~0$ so we have $\mathbb{G}^{M \dagger}_{21}=\mathbb{G}^{M}_{12}$. In addition, $\mathbb{T}_{i}(ik)$  is real and symmetric and in this case it can be written in the form 
$
\mathbb{T}_{i}=s_{i} \sqrt{s_{i}\mathbb{T}_{i}} \sqrt{s_{i}\mathbb{T}_{i}},
$
where $\sqrt{s_{i}\mathbb{T}_{i}}$ is the square root of the positive operator $s_{i} \mathbb{T}_{i}$.
We will not analyze convergence issues or the  self-adjointness of the presented operators (for those see \cite{hanson2013operator}), assuming that the appropriate  conditions are fulfilled in realistic systems \cite{rahi2010constraints}. 

As a result of the above, the interaction energy can be rewritten as \cite{romaniega2021repulsive}
\begin{equation}\label{eq:EnM}
	E_\text{int}=\frac{1}{2\pi}\int_0^\infty \mathrm{d}  k\,\text{Tr} \log(\mathbb{I}-s\, \mathbb{M}),
\end{equation}
where we have defined $s\equiv s_1 s_2$ and the nonnegative operator
\begin{equation} \label{eq:Proof}
	\mathbb{M}\equiv (\sqrt{s_2\mathbb{T}_2}\mathbb{G}^{M}_{21} \sqrt{s_1\mathbb{T}_1})^\dagger \sqrt{s_2\mathbb{T}_2}\mathbb{G}^{M}_{21} \sqrt{s_1\mathbb{T}_1}.
\end{equation}
In \cite{romaniega2021repulsive} we proved that the eigenvalues of $\mathbb{M}$ belong to $[0, 1)$ and that the sign of the interaction energy, including when both objects lie outside each other, is determined by
\begin{equation}\label{eq:SignEn}
	\textup{sgn}\,E_\textup{int}=-s=-s_1 s_2=- \textup{sgn}\left[(\varepsilon_{1}-\varepsilon_{M}\right)(\varepsilon_{2}-\varepsilon_{M})].
\end{equation}
In addition, for objects described by $\varepsilon_{i}$ and $\mu_{i}$, we showed that the relation $\textup{sgn}\, E_\textup{int}=-s_1s_2$ remains valid as long as the sign of the differential operator $\mathbb{V}_{i}(ik,\mathbf{x})$ in  Eq.~\eqref{eq:SchEM} is well-defined. This is determined by  Eq.~\eqref{eq:Signs} with an additional condition for the whole magnetodielectric \cite{rahi2009scattering}
\begin{equation}\label{eq:siMagnDiel}
	s_{i}=\pm 1 \quad \text{if} \  \varepsilon_{i}(ik, \textbf{x}) \gtrless \varepsilon_{M}(i k)\ \text{and}\ \mu_{i}(i k, \textbf{x})\lesseqgtr\mu_{M}(i k).
\end{equation}

\section{Inside scattering}\label{sec:InsideScatt}
\begin{figure}[h!]
	\centering
	\centering\includegraphics[width=0.4\textwidth]{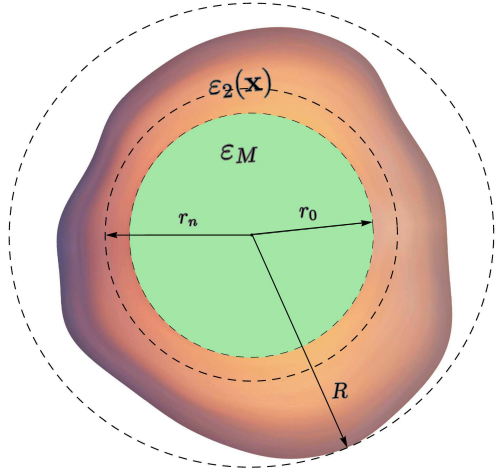}
	\caption{Cross-section view of the scattering object.  The radii of the smallest circumscribed sphere of the object and the spherical cavity wall are $R$ and $r_0$, respectively, being $r_n$ an intermediate radius.\label{fig:2}}  
\end{figure}

In this section we explore scattering processes within cavities. We are interested in how the probing wave can be scattered when it reaches the spherical cavity wall.  This unusual scattering setup, in which source and detector are inside the target, will be useful in the following section. 
In particular, we will focus on the variation of the scattering amplitude with the radius $r_0$ depicted in Fig.~\ref{fig:2}.
This will be determined employing the invariant imbedding technique \cite{sun2019invariant,johnson1988invariant}. This method, related to the variable phase approach of electromagnetic scattering,
is progressively reaching some importance in Casimir physics since it enables to compute efficiently  $\mathbb{T}_i$ for arbitrarily shaped objects \cite{forrow2012variable}. As we shall see, the derivation of this section holds for each fixed frequency. This will allow us to easily extend the result to systems at thermal equilibrium in Sec.~\ref{sec:Exten}. Therefore, we will omit this frequency dependence on the functions, except to indicate the analytic continuation to the imaginary frequency axis.
 
The starting point of the approach is the Lippmann-Schwinger equation \eqref{eq:LS} in position space
\begin{equation}\label{eq:LSIntegralEq}
\textbf{E}(\vecx)  = \textbf{E}_\text{in}(\vecx)- \int  \mathrm{d}\vecx'\mathbb{G}^{M}( \vecx,\vecx') \mathbb{V}_2(\omega, \vecx') \textbf{E}(\vecx'),
\end{equation}
 where $\textbf{E}_\text{in}$ is the incident wave. 
The dyadic Green's function defined in Eq.~\eqref{eq:DyadG} can be written as the sum of a transverse and a local contribution \cite{sun2019invariant}
\begin{equation}\label{eq:GTotal}
	\mathbb{G}^{M}(\vecx,\vecx')=\mathbb{G}^{M}_0(\vecx,\vecx')+\dfrac{\delta(\vecx-\vecx')}{\kappa^2}\vecu_\vecx \otimes \vecu_\vecx,
\end{equation}
where $\mathbb{G}^{M}_0(\vecx,\vecx')$ can be expanded as 
\begin{equation*}
\mathbb{G}^{M}_0(\vecx,\vecx')=-\kappa \sum_{\ell m} 
\left\{ \begin{array}{cc} \mathbb{Y}_{\ell m}(\Omega)\, \mathbb{H}^{\,}_\ell(i \kappa r) \mathbb{J}^\dagger_\ell(i \kappa r')\mathbb{Y}^\dagger_{\ell m}(\Omega')  & \, {\rm if}
\ \ r> r',\\[1ex]\mathbb{Y}_{\ell m}(\Omega)\, \mathbb{J}^{\,}_\ell(i \kappa r) \mathbb{H}^\dagger_\ell(i \kappa r')\mathbb{Y}^\dagger_{\ell m}(\Omega')  & \, {\rm if} \ \ r<r'.
	\end{array}\right.
\end{equation*}
We have defined $\omega= i k$, $\kappa\equiv  {k}{\sqrt{\varepsilon_{M}}}$ and
\begin{eqnarray}\label{eq:JandH}
	\mathbb{J}_\ell(i z)
	&\equiv& i^{\ell} \left(
	\begin{array}{cc}
		i_\ell(z) & 0 \\
		0 &  \frac{i }{z } \frac{\mathrm{d}\,  }{\mathrm{d}z } [z \, i_\ell(z )] \\[0.5ex]
		0 & \frac{i \sqrt{\ell (\ell+1)} }{z}  i_\ell(z) \\
	\end{array}
	\right)\!,\\[1ex]
	\mathbb{H}_\ell(i z)
	&\equiv& i^{\ell} \left(
	\begin{array}{cc}
		-  k_\ell(z) & 0 \\
		0 &  \frac{i }{z } \frac{\mathrm{d}\,  }{\mathrm{d} z }[z \, k_\ell(z )] \\[0.5ex]
		0 & \frac{i \sqrt{\ell (\ell+1)} }{z}  k_\ell(z)\\
	\end{array}
	\right),
\end{eqnarray}
being $i_\ell(z)$ and $k_\ell(z)$ the modified spherical Bessel functions 
\begin{equation}
i_\ell(z)\equiv \sqrt{\frac{\pi }{2 z}}\, I_{\ell+{1}/{2}}(z),\quad  k_\ell(z)\equiv \sqrt{\frac{2 }{\pi z}}\, K_{\ell+{1}/{2}}(z).
\end{equation}
The angular part is encoded in the square matrix $\mathbb{Y}_{\ell m}$ composed of vector spherical harmonics \cite{sun2019invariant}. In this case $\mathbb{Y}_{\ell m}(\Omega)\, \mathbb{H}^{\,}_\ell(i \kappa r)$ corresponds to an outgoing mode of the wave, where the first and second columns are  transverse electric (TE)  and transverse magnetic (TM) modes, respectively. Similarly, the first and second columns of $\mathbb{Y}_{\ell m}(\Omega)\, \mathbb{J}^{\,}_\ell(i \kappa r)$ represent regular TE and TM modes \cite{johnson1988invariant}. With these definitions we note that $\mathbb{G}^{M}_0(\vecx,\vecx')$ is subject to boundary conditions which correspond
to outgoing waves \cite{johnson1988invariant} and it can be rewritten as
\begin{equation*}
	\mathbb{G}^{M}_0(\vecx,\vecx')= \sum_{\ell m} \mathbb{Y}_{\ell m}(\Omega)\, \mathbb{G}^{\,}_\ell(r,r') \mathbb{Y}^\dagger_{\ell m}(\Omega'),
\end{equation*}
where
\begin{equation*}
	\mathbb{G}^{\,}_\ell(r,r')\equiv-\kappa \sum_{\ell m} 
	\left\{ \begin{array}{cc} \mathbb{H}^{\,}_\ell(i \kappa r) \mathbb{J}^\dagger_\ell(i \kappa r')  & \, {\rm if}
		\ \ r> r',\\[1ex]\mathbb{J}^{\,}_\ell(i \kappa r) \mathbb{H}^\dagger_\ell(i \kappa r')& \, {\rm if} \ \ r<r'.
	\end{array}\right.
\end{equation*}
The dielectric body is described by the potential of Eq.~\eqref{eq:Vi}
\begin{equation}\label{eq:V2}
\mathbb{V}_{2}(i k, \textbf{x}) = \mathbb{I} {V}_{2}(i k, \textbf{x}) =\mathbb{I} k^2 \left[\varepsilon_{2}(i k, \textbf{x})-\varepsilon_{M}(i k)\right],
\end{equation}
which from Eqs.~\eqref{eq:LSIntegralEq} and \eqref{eq:GTotal} satisfies
\begin{eqnarray}\label{eq:LSIntegralEq2}
\left(\mathbb{I}+\dfrac{V_2(i k, \vecx)}{\kappa^2}\vecu_\vecx \otimes \vecu_\vecx\right)\textbf{E}(\vecx)  
= \textbf{E}_\text{in}(\vecx)- \int  \mathrm{d}\vecx' {V}_2(i k,\vecx')\mathbb{G}^{M}_0(\vecx,\vecx')  \textbf{E}(\vecx').
\end{eqnarray}
Note that the spatial integral is over the volume of the second object $\text{supp}\,V_{2}$. We now define
\begin{equation*}
\textbf{E}^0(\vecx)\equiv\left(\mathbb{I}+\dfrac{V_2(i k, \vecx)}{\kappa^2}\vecu_\vecx \otimes \vecu_\vecx\right)\textbf{E}(\vecx)=\mathbb{D}^{-1}(\vecx)\textbf{E}(\vecx),
\end{equation*}
being $\mathbb{D}(\vecx)=\text{diag}(\varepsilon_M(i k) /\varepsilon_2(i k, \textbf{x}), 1,1)$ and $\textbf{E}^0(\vecx)=\textbf{E}(\vecx)$ if $r\notin[r_0, R]$. With this we rewrite Eq.~\eqref{eq:LSIntegralEq2} as
\begin{equation}\label{eq:E0}
\textbf{E}^0(\vecx)=\textbf{E}_\text{in}(\vecx)- \int  \mathrm{d}\vecx' {V}_{2}(i k, \textbf{x}') \mathbb{G}^{M}_0(\vecx,\vecx') \mathbb{D}(\vecx')\textbf{E}^0(\vecx') .
\end{equation}
Since the source of the scattering experiment is inside the cavity we  assume an outgoing incident wave 
\begin{equation}\label{eq:E02}
\textbf{E}^0_{\ell m}(\vecx)=\mathbb{Y}_{\ell m}(\Omega)\mathbb{H}^{\,}_{\ell}(i \kappa r)-\sum_{\ell' m'} \int  \mathrm{d}\vecx' {V}_{2}(i k, \textbf{x}') \mathbb{Y}_{\ell' m'}(\Omega)\, \mathbb{G}^{\,}_{\ell'}(r,r') \mathbb{Y}^\dagger_{\ell' m'}(\Omega')	 \mathbb{D}(\vecx')\textbf{E}^0_{\ell m}(\vecx') .
\end{equation}
In addition, since the detector is also inside we can write
\begin{eqnarray}\label{eq:LSInterior}
 	\textbf{E}^0_{\ell m}(\vecx)=\mathbb{Y}_{\ell m}(\Omega)\mathbb{H}^{\,}_{\ell}(i \kappa r)+ \sum_{\ell' m'}\mathbb{Y}_{\ell' m'}(\Omega)\mathbb{J}^{\,}_{\ell'}(i \kappa r)
\mathbb{T}^\text{int}_{\ell' m'\ell m}(ik),
\end{eqnarray}
being
\begin{equation}
\mathbb{T}^\text{int}_{\ell' m'\ell m}(ik)\equiv \kappa 
\int_{r_0}^R \mathrm{d} r'  r'^2 \mathbb{H}^\dagger_{\ell'}(i \kappa r') \int_{S^2} \mathrm{d}\Omega' {V}_{2}({ik}, \textbf{x}') \mathbb{Y}^\dagger_{\ell' m'}(\Omega') \mathbb{D}(\vecx')\textbf{E}^0_{\ell m}(\vecx').
\end{equation}
The physical interpretation of Eq.~\eqref{eq:LSInterior} is clear. The electric field detected is composed of the incident field and a combination of regular waves modulated by $\mathbb{T}^\text{int}(i k)$, where the four components of $\mathbb{T}^\text{int}_{\ell m\ell' m'}(ik)$ are determined by $\{\langle  \textbf{E}_{\ell m P}^\text{out}, \mathbb{{T}}\, \textbf{E}_{\ell' m' P'}^\text{out} \rangle\}_{P,\, P'\in \{\text{TE}, \text{TM}\}}$ \cite{sun2019invariant,rahi2009scattering}. 
It is instructive to compare the previous result with the regular scattering setup, in which source and detector are outside the target. In the latter, we have a regular wave and a scattered field composed of a combination of outgoing waves modulated by the usual transition matrix
\begin{eqnarray}\label{eq:LSExterior}
\textbf{E}^0_{\ell m}(\vecx)=\mathbb{Y}_{\ell m}(\Omega)\mathbb{J}^{\,}_{\ell}(i \kappa r)+ \sum_{\ell' m'}\mathbb{Y}_{\ell' m'}(\Omega)\mathbb{H}^{\,}_{\ell'}(i \kappa r)
\mathbb{T}^\text{ext}_{\ell' m'\ell m}(ik).
\end{eqnarray}
In both cases, in order to study  $\mathbb{T}(i k)$ it is useful to use Eq.~\eqref{eq:E02} for defining $\mathbb{F}_{\ell' m'\ell m}(r')$ by means of
	\begin{eqnarray*}
		\mathbb{F}_{\ell' m'\ell m}(r)&\equiv& r^2\int_{S^2} \mathrm{d}\Omega  \mathbb{Y}^\dagger_{\ell m}(\Omega){V}_{2}({ik}, \textbf{x}) \mathbb{D}(\vecx)\textbf{E}^0_{\ell' m'}(\vecx)\\
		&=& \mathbb{U}_{\ell' m'\ell m}(r) \mathbb{H}_{\ell'}(i \kappa r)-\sum_{\ell'' m''} \mathbb{U}_{\ell' m'\ell'' m''}(r)\int_{r_0}^R \mathrm{d} r' \mathbb{G}^{\,}_{\ell''}(r,r')\mathbb{F}_{\ell'' m''\ell' m'}(r')\nonumber,
	\end{eqnarray*}
being
\begin{equation}\label{eq:U}
	\mathbb{U}_{\ell m\ell' m'}(r)\equiv r^2\int_{S^2} \mathrm{d}\Omega {V}_{2}({ik}, \textbf{x}) \mathbb{Y}^\dagger_{\ell m}(\Omega) \mathbb{D}(\vecx)\mathbb{Y}_{\ell' m'}(\Omega).
\end{equation}
Note that $\mathbb{U}_{\ell m\ell' m'}(r)=0$ if $r\notin[r_0, R]$ and from Eq.~\eqref{eq:LSInterior} we have
\begin{equation}
	\mathbb{T}^\text{int}_{\ell' m'\ell m}(ik) = \kappa 
	\int_{r_0}^R \mathrm{d} r' \mathbb{H}^\dagger_{\ell'}(i \kappa r') {\mathbb{F}_{\ell' m'\ell m}(r')}. \label{eq:F}
\end{equation}
In order to use a more compact notation we can establish a unique correspondence between $\{\ell,\,m\}$ and a single index, which goes from one to infinity, and write \cite{sun2019invariant}
\begin{eqnarray}
\mathbb{F}(r)&=& \mathbb{U}(r) \mathbb{H}(i \kappa r)- \mathbb{U}(r)\int_{r_0}^R \mathrm{d} r' \mathbb{G}^{\,}(r,r')\mathbb{F}(r'),\label{eq:Fcompact} \\[0.5ex] \mathbb{T}^\text{int}(ik)&=& \kappa 
\int_{r_0}^R \mathrm{d} r' \mathbb{H}^\dagger(i \kappa r') {\mathbb{F}(r')}. \label{eq:Tcompact}
\end{eqnarray}
Now we can evaluate $\partial_{r_0}\mathbb{T}^\text{int}(i k)$ using the invariant imbedding technique  \cite{bi2013efficient}.
This method is based on considering the scattering by a cavity whose largest inscribed sphere radius is $r_n$ instead of $r_0$, as shown in Fig.~\ref{fig:2}. This lower limit can vary from $r_0$ to $R$ and if we start with $r_n$ close to $R$ we can keep imbedding spherical shells until the whole object is included. In the regular case this allows to transform a boundary condition problem into an initial condition one  \cite{sun2019invariant,johnson1988invariant}. Consequently, we define
\begin{eqnarray*}
\mathbb{U}(r_n;r)&\equiv&  \mathbb{U}(r) \theta(r-r_n),\\[0.5ex]
\mathbb{F}(r_n;r)&\equiv& \mathbb{U}(r_n;r) \mathbb{H}(i \kappa r)- \mathbb{U}(r_n;r)\int_{r_n}^{R} \mathrm{d} r' \mathbb{G}^{\,}(r,r')\mathbb{F}(r_n;r'), \\[0.5ex] \mathbb{T}^\text{int}(r_n; ik)&\equiv& \kappa 
\int_{r_n}^R \mathrm{d} r' \mathbb{H}^\dagger(i \kappa r') {\mathbb{F}(r_n; r')}. \label{eq:Trn}
\end{eqnarray*}
From these definitions we obtain
\begin{equation*}
\frac{\partial \mathbb{F}(r_n;r)}{\partial {r_n}}=-\kappa\mathbb{U}(r_n;r) \mathbb{H}(i \kappa r) \mathbb{J}^\dagger(i \kappa r_n)\mathbb{F}(r_n;r_n)-\mathbb{U}(r_n;r)\int_{r_n}^R \mathrm{d} r' \mathbb{G}(r,r') \frac{\partial \mathbb{F}(r_n;r')}{\partial {r_n}},
\end{equation*}
where the condition $r>r_n$ has been used. The previous equation can be written as
\begin{equation}\label{eq:dFrn}
\frac{\partial \mathbb{F}(r_n;r)}{\partial {r_n}}=-\kappa \mathbb{F}(r_n;r)\mathbb{J}^\dagger(i \kappa r_n)\mathbb{F}(r_n;r_n).
\end{equation}
Now, as can be seen from Eq.~\eqref{eq:Trn},
\begin{equation}
\frac{\partial \mathbb{T}^\text{int}(r_n; ik)}{\partial {r_n}}=-\kappa 
 \mathbb{H}^\dagger(i \kappa r_n) {\mathbb{F}(r_n; r_n)}+\kappa \int_{r_n}^R \mathrm{d} r'  \mathbb{H}^\dagger(i \kappa r') \frac{\partial \mathbb{F}(r_n;r')}{\partial {r_n}}.
\end{equation}
Noting that $\mathbb{G}(r_n,r')=-\kappa \mathbb{J}(i \kappa r_n)\mathbb{H}^\dagger(i \kappa r)$, we have
\begin{equation}\label{eq:Frn}
\mathbb{F}(r_n;r_n)= \mathbb{U}(r_n;r_n) [\mathbb{H}(i \kappa r_n)+ \mathbb{J}(i \kappa r_n)\mathbb{T}^\text{int}(r_n; ik)].
\end{equation}
Using Eqs.~\eqref{eq:dFrn}, \eqref{eq:Frn} and the fact that $\mathbb{T}_{i}(ik)$  is real and symmetric, we can finally write
\begin{equation}
\frac{\partial \mathbb{T}^\text{int}(r_n; ik)}{\partial {r_n}}=-\kappa \mathbb{W}^\dagger_\text{int}(r_n) \mathbb{U}(r_n;r_n) \mathbb{W}_\text{int}(r_n),
\end{equation}
where $\mathbb{W}_\text{int}(r_0)\equiv \mathbb{H}(i \kappa r_n)+\mathbb{J}(i \kappa r_n)\mathbb{T}^\text{int}(r_n; ik)$.  
In particular, for $r_n=r_0$ we have 
\begin{equation}\label{eq:dTinr0}
\frac{\partial \mathbb{T}^\text{int}( ik)}{\partial {r_0}}= -\kappa \mathbb{W}^\dagger_\text{int}(r_0) \mathbb{U}(r_0) \mathbb{W}_\text{int}(r_0).
\end{equation}
This nonlinear first-order differential equation is the basis of the numerical computation of $\mathbb{T}$ within the invariant imbedding approach. Again, it can be
compared with the one found in \cite{johnson1988invariant} for regular scattering. Adapted to imaginary frequencies with the definitions given in Eq.~\eqref{eq:JandH}, the derivative with respect to $R$ of $\mathbb{T}^\text{ext}( i k)$ in Eq.~\eqref{eq:LSExterior} is 
\begin{eqnarray}\label{eq:Calogero}
	\frac{\partial \mathbb{T}^\text{ext}( i k)}{\partial {R}}= \kappa \mathbb{W}_\text{ext}^\dagger(r_0) \mathbb{U}(R) \mathbb{W}_\text{ext}(r_0),
\end{eqnarray}
where $\mathbb{W}_\text{ext}(r_0)\equiv \mathbb{J}(i \kappa R)+\mathbb{H}(i \kappa R)\mathbb{T}^\text{ext}( ik)$.
To our knowledge, the latter was first proved in \cite{johnson1988invariant}, where the definition of $V(\omega,\vecx)$  differs from our definition by a global minus sign. Since the source is outside, the derivative is performed with respect to the smallest circumscribed sphere of the target. This equation for the transition  matrix is formally identical to the so-called Calogero equation of the quantum mechanical variable phase approach \cite{calogero1967variable}.

As we have seen in Sec.~\ref{sec:II}, in order to compute the Casimir-Lifshitz energy we only need on-shell matrix elements of $\mathbb{T}_i$. The latter are essentially   the scattering amplitudes $\mathbb{T}^\text{int}( ik)$ and $\mathbb{T}^\text{ext}( ik)$   {defined in Eqs.~\eqref{eq:LSInterior}  and \eqref{eq:LSExterior}. Note that for each body we employ a different complete set of free solutions of a given frequency. For instance, these components of $\mathbb{{T}}$ for the second object are determined by
	\begin{equation}
		\langle  \textbf{E}_{a}^\text{out},  \mathbb{{T}}_2 \textbf{E}_{a}^\text{out} \rangle =  \int \mathrm{d}\vecx \mathrm{d}\vecx' \textbf{E}_{a}^\text{out*}(i k, \vecx') \mathbb{{T}}_2(i k,\vecx',\vecx) \textbf{E}_{a}^\text{out}(i k, \vecx).
	\end{equation}
The singular behaviour of $\textbf{E}_{a}^\text{out*}(i k, \vecx)$ and  $\textbf{E}_{a}^\text{out*}(i k, \vecx')$ at the origin is avoided since $\mathbb{{T}}_2(i k, \vecx',\vecx)$ is nonzero only if both $\|\vecx\|$ and $\|\vecx'\| \in [r_0,R]$.
In this sense, if we consider the required components of $\mathbb{T}$ appearing in Eq.~\eqref{eq:ExpansionTGTG}, i.e., if we select the appropriate subspace determined by each complete set of solutions, and apply a Hellmann-Feynman argument \cite{feynman1939forces} using Eqs.~\eqref{eq:dTinr0} and \eqref{eq:Calogero} we can write 
	\begin{equation}\label{eq:DerT}
		s_1\partial_{R}\mathbb{T}_1>0, \quad s_2\partial_{r_{0}}\mathbb{T}_2<0.
	\end{equation}
	We have proved the first inequality explicitly for spherically symmetric bodies in \cite{romaniega2021repulsive}. In this case the problem is completely decoupled for $\{\ell, m, P\}$ so the square matrices $\mathbb{T}^\text{ext}_{\ell m\ell' m'}(ik)$ are diagonal. 

\section{Interaction pressure}\label{sec:IntPress}
There are cavity configurations in which the sum of the Casimir-Lifshitz forces  on each object equals zero. This is the case for systems invariant under mirror symmetry  with respect to the three spatial planes, such as the spherical one shown in Fig.~\ref{fig:1b}. 
\begin{figure}[h!]
	\centering
	\centering\includegraphics[width=0.4\textwidth]{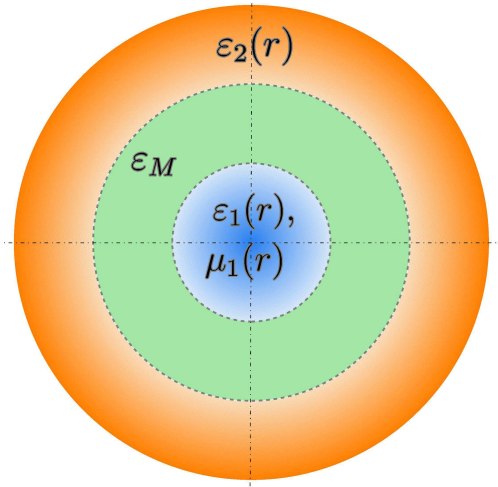}
	\caption{Cross-section view of a particular case of the system shown in Fig.~\ref{fig:1a}. For this spherical configuration  the sum of the Casimir-Lifshitz forces on each object equals zero.
		\label{fig:1b}}  
\end{figure}
However, the pressure acting on the surface of the bodies does not vanish. In this sense, we will study
the above-mentioned Casimir-Lifshitz interaction pressure acting on the cavity wall. The key idea is to obtain this pressure from the interaction energy given in Eq.~\eqref{eq:EnM}. This can be achieved introducing virtual variations of the radius $r_0$ shown in Fig.~\ref{fig:2} \cite{barton2004casimir}. Specifically, the mean value of the pressure on the wall  satisfies  \cite{li2019casimir}
\begin{eqnarray}\label{eq:MeanPress}
	\langle p_\text{int} \rangle \equiv  \dfrac{1}{4\pi}\int_{S^2} \mathrm{d}\Omega\, p_\text{int}(r_{0},\Omega) = -\dfrac{1}{4\pi r_{0}^{2}}\frac{\partial  E_\text{int}}{\partial  r_{0}}.
\end{eqnarray} 

In this way, we can straightforwardly find the sign of $\langle p_\textup{int}\rangle$. We simply note that only $\mathbb{T}_2$ depends on $r_{0}$, i.e., using the second equation of \eqref{eq:DerT}
we can write
$$
-\partial_{r_{0}} \mathbb{M}=-\left(\mathbb{G}^{M}_{21} \sqrt{s_1\mathbb{T}_1}\right)^\dagger \partial_{r_{0}}({s_2\mathbb{T}_2}) \left(\mathbb{G}^{M}_{21} \sqrt{s_1\mathbb{T}_1}\right)>0.
$$
In consequence, the derivative of the eigenvalues of $\mathbb{M}$ satisfy $\partial_{r_{0}} \lambda_\alpha~<~0$ \cite{feynman1939forces} so we have
\begin{equation}
\textup{sgn}[\partial_{r_{0}} \log(1-s\lambda_\alpha)]=s.
\end{equation}
Finally, from Eq.~\eqref{eq:EnM} and Lidskii's theorem we obtain $\text{sgn}\,\partial_{r_{0}}  E_\text{int}=s$, which can be written in terms of the pressure with Eq.~\eqref{eq:MeanPress} as
\begin{equation}\label{eq:SignPint}
	\textup{sgn}\langle p_\textup{int}\rangle= -s =- \textup{sgn}\left[(\varepsilon_{1}-\varepsilon_{M}\right)(\varepsilon_{2}-\varepsilon_{M})].
\end{equation}
As before, we have considered permittivity functions $\varepsilon(ik, \textbf{x})$ such that the sign of $\varepsilon_{i}(ik, \textbf{x})-\varepsilon_{M}(ik)$ is independent of $k$ and $\textbf{x}$, so we can write \begin{equation}
	s_i=\textup{sgn}(\varepsilon_{i}-\varepsilon_{M})=\textup{sgn}[\varepsilon_{i}(ik, \textbf{x})-\varepsilon_{M}(ik)].
\end{equation}

\subsection{Consistency tests and DLP configuration}\label{sec:TotalPressure}

	The previous result can be compared with particular  configurations  studied in the literature. First, for two concentric spherical shells  satisfying perfectly conducting boundary conditions a negative pressure, which tends to push the outer shell towards the inner shell, is found in \cite{teo2010casimir}. The latter is obtained using the zeta function regularization method and is consistent with Eq.~\eqref{eq:SignPint} since these idealized conditions arise in the limit of large permittivities. For the same boundary conditions this negative pressure is also obtained from Maxwell's stress tensor in the experimental setup suggested in \cite{brevik2005casimir}. 
	Furthermore, this pressure is computed within a quantum statistical approach for a system of two spherically shaped concentric dielectrics in \cite{hoye2001casimir}.  From the latter, it can be proved \cite{romaniega2021repulsive} that for the configuration shown in Fig.~\ref{fig:1b} with homogeneous dielectrics the sign of the  pressure $\textup{sgn}\, \langle p_\textup{int}\rangle$ equals to 
	\begin{equation*}
		\textup{sgn}\!\left\{\!\frac{-\left[\varepsilon_1(ik)-\varepsilon_{M}(ik)\right] \left[\varepsilon_2(ik)-\varepsilon_{M}(ik)\right] \ell(\ell+1) }{\left[\ell \varepsilon_1(ik)+\varepsilon_{M}(ik) (\ell+1) \right] \left[\ell \varepsilon_{M}(ik)+ \varepsilon_2(ik)(\ell+1)\right]}\!\right\},
	\end{equation*}
	which is in agreement with
	Eq.~\eqref{eq:SignPint}.
	
We conclude this section verifying Eq.~\eqref{eq:SignPint} against the DLP result. First of all, we assume the spherical configuration of Fig.~\ref{fig:1b}  and we take the limits $r_0, r_0'\to\infty$, where the difference between the radius of the sphere and the spherical shell $d\equiv r_0-r_0'>0$ is kept constant. With this we achieve the planar geometry \cite{cavero2021casimir}.  
Now, the sign of the force per unit area $\textup{sgn}\,F_\textup{int}$ can be found from Eq.~\eqref{eq:SignPint}. We first note that  
\begin{equation}
\textup{sgn}\,F_\textup{int} =-\textup{sgn}\,\partial_{d}E_\text{int}=-\textup{sgn}\, \partial_{r_0}E_\text{int}.
\end{equation}
Bearing in mind Eqs.~\eqref{eq:MeanPress} and \eqref{eq:SignPint} we can finally write
\begin{equation}\label{eq:SignFint}
\textup{sgn}\,F_\textup{int} =- \textup{sgn}\left[(\varepsilon_{1}-\varepsilon_{M}\right)(\varepsilon_{2}-\varepsilon_{M})].
\end{equation}
This is the DLP result cited in Sec.~\ref{sec:I}.
As we have mentioned, both the repulsive and the attractive interactions resulting from this configuration have been confirmed experimentally \cite{munday2009measured}. Furthermore, we have proved that the DLP result in \cite{dzyaloshinskii1961general} can be extended to  inhomogeneous slabs as long as Eq.~\eqref{eq:Signs} holds.
The latter is in agreement with the result we found in \cite{romaniega2021repulsive}. In the latter, the DLP result was recovered in a similar way but computing the pressure on a sphere enclosed within an arbitrarily shaped cavity. For this configuration we obtained
\begin{equation}
	\textup{sgn}\,E_\textup{int}=\textup{sgn}\,F_\textup{int}=-\textup{sgn}\, \langle p_\textup{int}\rangle,
\end{equation}
where $\langle p_\textup{int}\rangle$ here denotes the mean pressure on the inner sphere.

\section{Extensions}\label{sec:Exten}
In the previous sections, we have determined the signs of the interaction energy and pressure, finding that both depend on the difference between the permittivities of the bodies and the medium in the same way
\begin{equation}\label{eq:MainRes}
\textup{sgn}\,E_\textup{int} = \textup{sgn}\,\langle p_\text{int} \rangle = -\textup{sgn}\left[(\varepsilon_{1}-\varepsilon_{M}\right)(\varepsilon_{2}-\varepsilon_{M})].
\end{equation}
In this final section we study two generalizations of the system considered so far, showing that these results remain valid with some minor changes.

Firstly, we can consider a magnetodielectric inner object.
The derivation of Sec.~\ref{sec:InsideScatt} holds for a homogeneous background and a nonmagnetic scattering object. However, in order to determine $\text{sgn}\, \langle p_\textup{int}\rangle$ we have only assumed a well-defined sign of  $\mathbb{T}_1(ik)$.  Therefore, if the whole  body is described by permittivity and permeability functions such that
\begin{equation*}
s_{1}=\pm 1 \quad \text{if} \  \varepsilon_{1}(ik, \textbf{x}) \gtrless \varepsilon_{M}(i k)\ \text{and}\ \mu_{1}(i k, \textbf{x})~\lesseqgtr~\mu_{M}(i k)
\end{equation*}
the relation $\textup{sgn}\,\langle p_\text{int} \rangle =-s_1 s_2$ remains valid. 

We can also extend  the aforementioned results to quantum systems at thermal equilibrium. In analogy to the interaction energy, the free energy $\mathcal{F}_\text{int}$  satisfies \cite{milton2001casimir}
\begin{equation}\label{eq:PFint}
\langle p_\text{int} (T) \rangle =-\frac{1}{4\pi r_{0}^{2}}\,\frac{\partial  \mathcal{F}_\text{int}(T)}{\partial  r_{0}}.
\end{equation}
In this case we can compute $\mathcal{F}_\text{int}(T)$  replacing the integral in $E_\textup{int}$ by a sum over the Matsubara frequencies $k_{n}~=~2\pi k_\text{B} {n} T$, where the zero mode is weighted by $1/2$  \cite{bordag2009advances}
\begin{equation}
\mathcal{F}_\text{int}= k_B T \sum_{n=0}^{\infty}{}^\prime \ \text{Tr} \log\left[\mathbb{I}-s\, \mathbb{M}(i k_{n})\right].
\end{equation} 
Since the proofs of \eqref{eq:MainRes} apply to each fixed frequency, they will also hold for systems at thermal equilibrium. 
\section{Conclusions}\label{conclusions}

In this paper, we have extended previous work on the sign of the interaction pressure on spherical surfaces.
Again, we find that the sign of the interaction pressure and energy changes as follows:
\begin{equation}\label{eq:MainRes2}
\textup{sgn}\,E_\textup{int} = \textup{sgn}\,\langle p_\text{int} \rangle = -\textup{sgn}\left[(\varepsilon_{1}-\varepsilon_{M}\right)(\varepsilon_{2}-\varepsilon_{M})]=-s.
\end{equation}
This behavior was first found by DLP  in \cite{dzyaloshinskii1961general}, where they extended
Casimir’s
formulation for ideal metal plates in vacuum to dielectric
materials.
Our result is independent of the geometry of the objects as long as the assumption \eqref{eq:Signs} on the sign of the potentials holds.  Since the proof applies to each frequency independently, the extension to systems at thermal equilibrium is almost immediate. Expressing the energy in terms of transition operator also enables us to generalize the previous result to a magnetodielectric inner object and to recover the DLP result, generalized to inhomogeneous slabs, as a limiting case.

In order to determine the attractive or repulsive character of the fluctuation-induced pressure acting on the cavity wall, the self-energy contribution should also be considered, as we did in \cite{romaniega2021repulsive}. Nevertheless, the interaction term itself already leads to theoretical and experimental implications. We can illustrate some of them considering the spherical system of Fig.~\ref{fig:1b}. From \cite{romaniega2021repulsive} and Eq.~\eqref{eq:MainRes2} we conclude that the interaction term
always gives rise to opposite behaviours on both  spherical surfaces, i.e., 	$\textup{sgn}\, \langle p_\textup{int}\rangle = \pm s$ on the sphere and cavity wall, respectively.
For instance, when the permittivities of the objects and medium  satisfy $\varepsilon_{i}<\varepsilon_{M}<\varepsilon_{j}$, $i,j\in\{1,2\}$, we obtain a positive interaction energy and  the pressure acting on these spherical surfaces tends to repel each other. That is to say, the interaction terms tends to contract the inner surface and expand the outer one, which is consistent with the experimental verification of the above-mentioned repulsion between material bodies \cite{munday2009measured}. For values of the permittivities resulting in a negative interaction energy,  the pressure tends to push the inner surface to the outer one, and vice versa. For instance, this attractive behaviour necessarily arises when the objects are immersed in vacuum. From the theorem on the stability by Casimir–Lifshitz forces \cite{rahi2010constraints}, which is also governed by the sign of $(\varepsilon_{1}-\varepsilon_{M})(\varepsilon_{2}-\varepsilon_{M})$, we know that in the previous case any equilibrium position of the two objects
subject to such forces is unstable.

To end with, we note that although the configurations considered in this paper and in \cite{romaniega2021repulsive} are different, in both cases the conclusions are based on the study of the scattering amplitudes of each body. The main difference is the scattering amplitudes needed: the regular ones in \cite{romaniega2021repulsive} and the ones associated with an unusual scattering setup in which source and detector are inside the target in this paper. For the latter, we have derived the suitable expressions using the invariant imbedding technique.

\section*{Acknowledgments}
I am grateful to A. Romaniega, L. M. Nieto, I. Cavero\,-Pel\'aez   and J. M. Mu\~noz-Casta\~neda for the useful suggestions.
This work was  supported by the FPU fellowship program (FPU17/01475) and the  Junta de Castilla y Le\'on and FEDER  (Project BU229\-P18).

\appendix

\section{Inside and outside scattering}
\label{sec:Appendices}

Similar to the proof in Sec.~\ref{sec:InsideScatt}, we can study the variation of the scattering amplitudes with respect to a parameter defining the body, such as the radius of the cavity wall. In this  appendix we perform a similar derivation for two different configurations. Both of them are determined by the position of  the source and the detector in the scattering experiment. As we shall see, in these two cases the derivative can not be written as in Eqs.~\eqref{eq:dTinr0} and \eqref{eq:Calogero}.

\subsection{Detector outside, source inside}
Similar to the proof of Sec.~\ref{sec:InsideScatt}, the starting point is the Lippmann-Schwinger equation \eqref{eq:E0} 
\begin{equation}\label{eq:E0ei}
\textbf{E}^0(\vecx)=\textbf{E}_\text{in}(\vecx)- \int  \mathrm{d}\vecx' {V}_{2}(i k, \textbf{x}') \mathbb{G}^{M}_0(\vecx,\vecx') \mathbb{D}(\vecx')\textbf{E}^0(\vecx') .
\end{equation}
We are interested in a scattering process in which the source is inside the cavity but the detector is outside.
Consequently, we  assume an outgoing incident wave
\begin{equation}\label{eq:E02ei}
\textbf{E}^0_{\ell m}(\vecx)=\mathbb{Y}_{\ell m}(\Omega)\mathbb{H}^{\,}_{\ell}(i \kappa r)-\sum_{\ell' m'} \int  \mathrm{d}\vecx' {V}_{2}(i k, \textbf{x}') \mathbb{Y}_{\ell' m'}(\Omega)\, \mathbb{G}^{\,}_{\ell'}(r,r') \mathbb{Y}^\dagger_{\ell' m'}(\Omega')	 \mathbb{D}(\vecx')\textbf{E}^0_{\ell m}(\vecx') .
\end{equation}
In addition, since the detector is outside,
\begin{eqnarray}\label{eq:A3}
	\textbf{E}^0_{\ell m}(\vecx)=\mathbb{Y}_{\ell m}(\Omega)\mathbb{H}^{\,}_{\ell}(i \kappa r)+ \sum_{\ell' m'}\mathbb{Y}_{\ell' m'}(\Omega)\mathbb{H}^{\,}_{\ell'}(i \kappa r)
 \mathbb{T}^\text{ei}_{\ell' m'\ell m}(ik),
\end{eqnarray}
being
\begin{equation}
 \mathbb{T}^\text{ei}_{\ell' m'\ell m}(ik)\equiv \kappa 
\int_{r_0}^R \mathrm{d} r'  r'^2 \mathbb{J}^\dagger_{\ell'}(i \kappa r') \int_{S^2} \mathrm{d}\Omega' {V}_{2}({ik}, \textbf{x}') \mathbb{Y}^\dagger_{\ell' m'}(\Omega') \mathbb{D}(\vecx')\textbf{E}^0_{\ell m}(\vecx').
\end{equation}
Equation \eqref{eq:A3} is analogous to Eq.~(4.12) in \cite{rahi2009scattering}. The electric field detected is composed of the incident field and a combination of outgoing waves modulated by 
\begin{equation}
 \mathbb{T}^\text{ei}_{\ell' m'\ell m}(ik)= \kappa 
\int_{r_0}^R \mathrm{d} r' \mathbb{J}^\dagger_{\ell'}(i \kappa r') {\mathbb{F}_{\ell' m'\ell m}(r')} \label{eq:Fei}.
\end{equation}
If we insert Eq.~\eqref{eq:E02ei} in $\mathbb{F}_{\ell m\ell' m'}(r)$  we obtain the following relation:
\begin{equation}
\mathbb{F}_{\ell' m'\ell m}(r)= \mathbb{U}_{\ell' m'\ell m}(r) \mathbb{H}_{\ell'}(i \kappa r)-\sum_{\ell'' m''} \mathbb{U}_{\ell' m'\ell'' m''}(r)\int_{r_0}^R \mathrm{d} r' \mathbb{G}^{\,}_{\ell''}(r,r')\mathbb{F}_{\ell'' m''\ell' m'}(r')\nonumber,
\end{equation}
where $\mathbb{U}_{\ell m\ell' m'}(r)$ has been defined in Eq.~\eqref{eq:U}.
As in Sec.~\ref{sec:InsideScatt}, these results can be written in compact notation \cite{sun2019invariant}
\begin{eqnarray}
\mathbb{F}(r)&=& \mathbb{U}(r) \mathbb{H}(i \kappa r)- \mathbb{U}(r)\int_{r_0}^R \mathrm{d} r' \mathbb{G}^{\,}(r,r')\mathbb{F}(r'),\label{eq:Fcompactei} \\[0.5ex]  \mathbb{T}^\text{ei}(ik)&=& \kappa 
\int_{r_0}^R \mathrm{d} r' \mathbb{J}^\dagger(i \kappa r') {\mathbb{F}(r')}, \label{eq:Tcompactei}
\end{eqnarray}
and in order to compute $\partial_{r_0} \mathbb{T}^\text{ei}(i k)$ in the present case we define
\begin{eqnarray}
\mathbb{U}(r_n;r)&\equiv&  \mathbb{U}(r) \theta(r-r_n),\\[0.5ex]
\mathbb{F}(r_n;r)&\equiv& \mathbb{U}(r_n;r) \mathbb{H}(i \kappa r)- \mathbb{U}(r_n;r)\int_{r_n}^R \mathrm{d} r' \mathbb{G}^{\,}(r,r')\mathbb{F}(r_n;r'), \\[0.5ex]  
\mathbb{T}^\text{ei}(r_n; ik)&\equiv& \kappa 
\int_{r_n}^R \mathrm{d} r' \mathbb{J}^\dagger(i \kappa r') {\mathbb{F}(r_n; r')}. \label{eq:Trnei}
\end{eqnarray}
Now, evaluating $\partial_{r_n}\mathbb{F}(r_n;r)$ we obtain
\begin{eqnarray*}
	\frac{\partial \mathbb{F}(r_n;r)}{\partial {r_n}}&=&\mathbb{U}(r_n;r) \mathbb{G}(r,r_n)\mathbb{F}(r_n;r_n)-\mathbb{U}(r_n;r)\int_{r_n}^R \mathrm{d} r' \mathbb{G}(r,r') \frac{\partial \mathbb{F}(r_n;r')}{\partial {r_n}} \\
	&=&-\kappa\mathbb{U}(r_n;r) \mathbb{H}(i \kappa r) \mathbb{J}^\dagger(i \kappa r_n)\mathbb{F}(r_n;r_n)-\mathbb{U}(r_n;r)\int_{r_n}^R \mathrm{d} r' \mathbb{G}(r,r') \frac{\partial \mathbb{F}(r_n;r')}{\partial {r_n}},
\end{eqnarray*}
where the condition $r>r_n$ has been used. The previous equation can be written as
\begin{equation}\label{eq:dFrnei}
\frac{\partial \mathbb{F}(r_n;r)}{\partial {r_n}}=-\kappa \mathbb{F}(r_n;r)\mathbb{J}^\dagger(i \kappa r_n)\mathbb{F}(r_n;r_n).
\end{equation}
Now, as can be seen from Eq.~\eqref{eq:Trnei}
\begin{equation}
\frac{\mathrm{d}  \mathbb{T}^\text{ei}(r_n; ik)}{\mathrm{d} {r_n}}=-\kappa 
\mathbb{J}^\dagger(i \kappa r_n) {\mathbb{F}(r_n; r_n)}+\kappa \int_{r_n}^R \mathrm{d} r'  \mathbb{J}^\dagger(i \kappa r') \frac{\partial \mathbb{F}(r_n;r')}{\partial {r_n}},
\end{equation} 
and noting that $\mathbb{G}(r_n,r')=-\kappa \mathbb{J}(i \kappa r_n)\mathbb{H}^\dagger(i \kappa r) $ we have
\begin{equation}\label{eq:Frnei}
\mathbb{F}(r_n;r_n)= \mathbb{U}(r_n;r_n) [\mathbb{H}(i \kappa r_n)+ \mathbb{J}(i \kappa r_n) \mathbb{T}^\text{ei}(r_n; ik)].
\end{equation}
Using Eqs.~\eqref{eq:dFrnei} and \eqref{eq:Frnei} we can finally write
\begin{equation}
	\frac{\mathrm{d}  \mathbb{T}^\text{ei}(r_n; ik)}{\mathrm{d} {r_n}} = -\kappa [\mathbb{J}^\dagger(i \kappa r_n)+{\mathbb{T}^\text{ei}}(r_n; ik)\mathbb{J}^\dagger(i \kappa r_n) ] \mathbb{F}(r_n;r_n).
\end{equation}
Note that the structure of this derivative and the one in Eq.~\eqref{eq:dTinr0} are entirely different. 
 In this case the scattering amplitude is essentially $\langle \textbf{E}^\text{reg}_a,\mathbb{T} \textbf{E}^\text{out}_b\rangle$ and the sign is not necessarily well-defined since the derivative cannot be written as $-\kappa \mathbb{W}^\dagger_\text{ei}(r_n) \mathbb{U}(r_n) \mathbb{W}_\text{ei}(r_n)$.

\subsection{Detector inside, source outside}
Similar to the proof in Sec.~\ref{sec:InsideScatt}, the starting point is the Lippmann-Schwinger equation \eqref{eq:E0} 
\begin{equation}\label{eq:E0ie}
\textbf{E}^0(\vecx)=\textbf{E}_\text{in}(\vecx)- \int  \mathrm{d}\vecx' {V}_{2}(i k, \textbf{x}') \mathbb{G}^{M}_0(\vecx,\vecx') \mathbb{D}(\vecx')\textbf{E}^0(\vecx') .
\end{equation}
We are interested in a scattering process in which the source is outside the cavity being the detector inside.
Consequently, we  assume a regular incident wave
\begin{equation}\label{eq:E02ie}
\textbf{E}^0_{\ell m}(\vecx)=\mathbb{Y}_{\ell m}(\Omega)\mathbb{J}^{\,}_{\ell}(i \kappa r)-\sum_{\ell' m'} \int  \mathrm{d}\vecx' {V}_{2}(i k, \textbf{x}') \mathbb{Y}_{\ell' m'}(\Omega)\, \mathbb{G}^{\,}_{\ell'}(r,r') \mathbb{Y}^\dagger_{\ell' m'}(\Omega')	 \mathbb{D}(\vecx')\textbf{E}^0_{\ell m}(\vecx') .
\end{equation}
In addition, since the detector is inside,
\begin{eqnarray}\label{eq:A17}
\textbf{E}^0_{\ell m}(\vecx)=\mathbb{Y}_{\ell m}(\Omega)\mathbb{J}^{\,}_{\ell}(i \kappa r)+ \sum_{\ell' m'}\mathbb{Y}_{\ell' m'}(\Omega)\mathbb{J}^{\,}_{\ell'}(i \kappa r)
\mathbb{T}^\text{ie}_{\ell' m'\ell m}(ik),
\end{eqnarray}
being
\begin{equation}
\mathbb{T}^\text{ie}_{\ell' m'\ell m}(ik)\equiv \kappa 
\int_{r_0}^R \mathrm{d} r'  r'^2 \mathbb{H}^\dagger_{\ell'}(i \kappa r') \int_{S^2} \mathrm{d}\Omega' {V}_{2}({ik}, \textbf{x}') \mathbb{Y}^\dagger_{\ell' m'}(\Omega') \mathbb{D}(\vecx')\textbf{E}^0_{\ell m}(\vecx').
\end{equation}
Equation \eqref{eq:A17} is analogous to Eq.~(4.10) in \cite{rahi2009scattering}. The electric field detected is composed of the incident field  and a combination of regular waves modulated by
\begin{equation}
\mathbb{T}^\text{ie}_{\ell' m'\ell m}(ik)= \kappa 
\int_{r_0}^R \mathrm{d} r' \mathbb{H}^\dagger_{\ell'}(i \kappa r') {\mathbb{F}_{\ell' m'\ell m}(r')} \label{eq:Fie}.
\end{equation}
If we insert Eq.~\eqref{eq:E02ie} in $\mathbb{F}_{\ell m\ell' m'}(r)$ we obtain the following relation:
\begin{equation}
\mathbb{F}_{\ell' m'\ell m}(r)= \mathbb{U}_{\ell' m'\ell m}(r) \mathbb{J}_{\ell'}(i \kappa r)-\sum_{\ell'' m''} \mathbb{U}_{\ell' m'\ell'' m''}(r)\int_{r_0}^R \mathrm{d} r' \mathbb{G}^{\,}_{\ell''}(r,r')\mathbb{F}_{\ell'' m''\ell' m'}(r')\nonumber,
\end{equation}
where $\mathbb{U}_{\ell m\ell' m'}(r)$ has been defined in Eq.~\eqref{eq:U}.
As in Sec.~\ref{sec:InsideScatt}, these results can be written in compact notation \cite{sun2019invariant}
\begin{eqnarray}
\mathbb{F}(r)&=& \mathbb{U}(r) \mathbb{J}(i \kappa r)- \mathbb{U}(r)\int_{r_0}^R \mathrm{d} r' \mathbb{G}^{\,}(r,r')\mathbb{F}(r'),\label{eq:Fcompactie} \\[0.5ex]  \mathbb{T}^\text{ie}(ik)&=& \kappa 
\int_{r_0}^R \mathrm{d} r' \mathbb{H}^\dagger(i \kappa r') {\mathbb{F}(r')}, \label{eq:Tcompactie}
\end{eqnarray}
and in order to compute $\partial_{r_0} \mathbb{T}^\text{ie}(i k)$ in the present case we define
\begin{eqnarray}
\mathbb{U}(r_n;r)&\equiv&  \mathbb{U}(r) \theta(r_n-r),\\[0.5ex]
\mathbb{F}(r_n;r)&\equiv& \mathbb{U}(r_n;r) \mathbb{J}(i \kappa r)- \mathbb{U}(r_n;r)\int_{r_0}^{r_n} \mathrm{d} r' \mathbb{G}^{\,}(r,r')\mathbb{F}(r_n;r'), \\[0.5ex]  
\mathbb{T}^\text{ie}(r_n; ik)&\equiv& \kappa 
\int_{r_0}^{r_n} \mathrm{d} r' \mathbb{H}^\dagger(i \kappa r') {\mathbb{F}(r_n; r')}. \label{eq:Trnie}
\end{eqnarray}
Note that we are considering the derivative with respect to $r_n$, being the new body defined by the potential $V_2(i k) \theta (r_n-r)$ and not $V_2(i k) \theta (r-r_n)$. Now, evaluating $\partial_{r_n}\mathbb{F}(r_n;r)$ we obtain
\begin{eqnarray*}
	\frac{\partial \mathbb{F}(r_n;r)}{\partial {r_n}}&=&-\mathbb{U}(r_n;r) \mathbb{G}(r,r_n)\mathbb{F}(r_n;r_n)-\mathbb{U}(r_n;r)\int_{r_0}^{r_n} \mathrm{d} r' \mathbb{G}(r,r') \frac{\partial \mathbb{F}(r_n;r')}{\partial {r_n}} \\
	&=&\kappa\mathbb{U}(r_n;r) \mathbb{J}(i \kappa r) \mathbb{H}^\dagger(i \kappa r_n)\mathbb{F}(r_n;r_n)-\mathbb{U}(r_n;r)\int_{r_0}^{r_n} \mathrm{d} r' \mathbb{G}(r,r') \frac{\partial \mathbb{F}(r_n;r')}{\partial {r_n}},
\end{eqnarray*}
where the condition $r<r_n$ has been used. The previous equation can be written as
\begin{equation}\label{eq:dFrnie}
\frac{\partial \mathbb{F}(r_n;r)}{\partial {r_n}}=\kappa \mathbb{F}(r_n;r)\mathbb{H}^\dagger(i \kappa r_n)\mathbb{F}(r_n;r_n).
\end{equation}
Now, as can be seen from Eq.~\eqref{eq:Trnie}
\begin{equation}
\frac{\mathrm{d}  \mathbb{T}^\text{ie}(r_n; ik)}{\mathrm{d} {r_n}}=\kappa 
\mathbb{H}^\dagger(i \kappa r_n) {\mathbb{F}(r_n; r_n)}+\kappa \int_{r_0}^{r_n} \mathrm{d} r'  \mathbb{H}^\dagger(i \kappa r') \frac{\partial \mathbb{F}(r_n;r')}{\partial {r_n}},
\end{equation}
and noting that $\mathbb{G}(r_n,r')=-\kappa \mathbb{H}(i \kappa r_n)\mathbb{J}^\dagger(i \kappa r) $ we have
\begin{equation}\label{eq:Frnie}
\mathbb{F}(r_n;r_n)= \mathbb{U}(r_n;r_n) [\mathbb{J}(i \kappa r_n)+ \mathbb{H}(i \kappa r_n) \mathbb{T}^\text{ie}(r_n; ik)].
\end{equation}
Using Eqs.~\eqref{eq:dFrnie} and \eqref{eq:Frnie} we can finally write
\begin{equation}
\frac{\mathrm{d}  \mathbb{T}^\text{ie}(r_n; ik)}{\mathrm{d} {r_n}} = \kappa [\mathbb{H}^\dagger(i \kappa r_n)+{\mathbb{T}^\text{ie}}(r_n; ik)\mathbb{H}^\dagger(i \kappa r_n) ] \mathbb{F}(r_n;r_n).
\end{equation}
As in the previous section, the structure of this derivative and the one in Eq.~\eqref{eq:dTinr0} are different. In this case the scattering amplitude is essentially $\langle \textbf{E}^\text{out}_a,\mathbb{T} \textbf{E}^\text{reg}_b\rangle$ and the sign is also not necessarily well-defined since the derivative cannot be written as $-\kappa \mathbb{W}^\dagger_\text{ie}(r_n) \mathbb{U}(r_n) \mathbb{W}_\text{ie}(r_n)$. Indeed, we have $\langle \textbf{E}^\text{out}_a,\mathbb{T} \textbf{E}^\text{reg}_b\rangle^* = \langle \textbf{E}^\text{reg}_a,\mathbb{T} \textbf{E}^\text{out}_b\rangle$.

\end{document}